\documentclass[aps,pra,twocolumn,showpacs,floatfix]{revtex4-2}
\usepackage{graphicx}
\usepackage{amsmath}
\usepackage{amssymb}
\usepackage{color}
\usepackage[separate-uncertainty=true]{siunitx}
\usepackage{lipsum}

\newcommand{\ket}[1]{\left| #1\right\rangle}

\newcommand{\Hetwo}[0]{$\mathrm{He}_2^*$}

\newcommand{\Hefour}[0]{$^4\mathrm{He}$}
\newcommand{\HeII}{He\,II}
\newcommand{\Astate}[0]{$a^3\Sigma_u^+$}
\newcommand{\Bstate}[0]{$b^3\Pi_g$}

\newcommand{\Dstate}[0]{$d^3\Sigma_u^+$}
\newcommand{\LDlif}[1]{$LD_\textsc{lif}#1$}
\newcommand{\E}{{\cal E}}

\begin{document}

\title{Dynamics of molecular rotors in bulk superfluid helium}

\author{Alexander~A.~Milner$^{1}$, V.~A.~Apkarian$^{2}$ , and Valery~Milner$^{1}$}

\affiliation{$^{1}$Department of  Physics \& Astronomy, The University of British Columbia, Vancouver, Canada}
\affiliation{$^{2}$Department of Chemistry, University of California, Irvine, California 92697, United States}

\date{\today}

\begin{abstract}
Molecules immersed in liquid helium are excellent probes of superfluidity. Their electronic, vibrational and rotational dynamics provide valuable clues about the superfluid at the nanoscale. Here we report on the experimental study of the laser-induced rotation of helium dimers inside the superfluid \Hefour{} bath at variable temperature. The coherent rotational dynamics of \Hetwo{} is initiated in a controlled way by ultrashort laser pulses, and tracked by means of time-resolved laser-induced fluorescence. We detect the decay of rotational coherence on the nanosecond timescale and investigate the effects of temperature on the decoherence rate. The observed temperature dependence suggests a non-equilibrium evolution of the quantum bath, accompanied by the emission of the wave of second sound. The method offers new ways of studying superfluidity with molecular nano-probes under variable thermodynamic conditions.
\end{abstract}
\maketitle

Elementary excitations in liquid helium (LHe) and its superfluid phase (\HeII) have been studied predominantly with neutron scattering \cite{Glyde2017}, as well as by observing the dynamics of embedded atoms and molecules \cite{Toennies1998}. Due to the vanishingly small solubility of impurities in LHe, the use of molecular probes has been largely limited to studies in helium nanodroplets that can be doped by injection of foreign species in pick-up cells \cite{Toennies2004, Stienkemeier2006, Slenczka2022}. A wealth of information has been extracted from such studies about the coupling between the molecular electronic, vibrational and rotational degrees of freedom and the quantum bath, be it through frequency \cite{Grebenev1998, Choi2006, Lehnig2009} or time domain \cite{Nielsen2022} measurements. As the microscopic analog of the Andronikashvili experiment, which used a torsion balance to verify the phenomenological two-fluid model of \HeII, molecular rotors have been most informative: The change in the moment of inertia and centrifugal distortion constant of an embedded molecule serves as a gauge of the dragged normal fraction, and nearly free rotation is taken as the signature of a frictionless superfluid bath \cite{Shepperson2017, Chatterley2020, Cherepanov2021, Qiang2022}.

Despite their elegance, nanodroplets suffer from a serious limitation: their thermodynamic state is fixed to a single point on the temperature-pressure $(T,P)$ plane because of the evaporative cooling used in their production. Yet to investigate the inherently macroscopic two-fluid model of \HeII, it is essential to carry out measurements as a function of thermodynamic variables. This can be accomplished by resorting to the helium dimers in the lowest metastable triplet state (\Astate{}), known as \Hetwo{} excimers, as liquid helium's native molecular probes \cite{Surko1968, Dennis1969, Hill1971}. With a lifetime on the order of a few seconds \cite{Keto1974, Benderskii1999, McKinsey2003}, the excimers are ideally suited for time-resolved studies of their interaction with the quantum environment.

Similarly to solvated electrons, \Hetwo{} excimers form in $\sim\SI{14}{\AA}$-diameter cavities (or ``bubbles'') that expel the superfluid around the molecule \cite{Dennis1969, Eloranta2001, Eloranta2002}. Electronic transitions in \Hetwo{} have been used to drive damped bubble oscillations, whose dependence on temperature and pressure was shown to track the normal fraction, thus establishing that the two-fluid model extends down to the molecular scale \cite{Benderskii2002}. Rotational lines in the fluorescence spectra, albeit unresolved but with the envelope similar to that in the gas phase, indicated free rotation of \Hetwo{} inside the bubble \cite{Dennis1969, Hill1971}. However, large inhomogeneous broadening \cite{Eltsov1995, Rellergert2008} due to bubble shape fluctuations \cite{Guo2020} prohibited the spectroscopic analysis of the excimer's rotational dynamics. The observed slow time dependence of the broadened absorption lineshape indicated the characteristic timescale for the rotational cooling of a few milliseconds \cite{Eltsov1995, Eltsov1998}, but offered no information on the (potentially much faster) decay of rotational coherence and, therefore, on the finer details of the molecular interaction with \HeII{}. With no direct access to molecular rotation in \textit{bulk superfluid}, the microscopic Andronikashvili experiment under controlled thermodynamic conditions remained unrealized.

In the time-domain study presented here, we prepare coherent rotational wave packets in \Hetwo{}, and investigate their decoherence with femtosecond resolution in the superfluid quantum bath at variable temperature. After producing $a$-state excimers with intense pump pulses \cite{Benderskii1999, McKinsey2003}, we excite molecular rotation by a linearly polarized fs ``kick'' pulse, and then follow it in time with a delayed probe (see Supplemental Material for details \cite{Supplement}). Two-photon probe absorption promotes the molecule to a fluorescent $d$ state (\Dstate{}), which decays to \Bstate{} by emitting a photon at $\approx \SI{640}{nm}$ \cite{Benderskii1999, Rellergert2008, Guo2014, Gao2015}. Owing to the anisotropic absorption cross-section, the difference between the laser-induced fluorescence (LIF) signals corresponding to two orthogonal probe polarizations (known, and hereafter referred to, as ``linear dichroism'' \LDlif{} \cite{Supplement}) reflects the ensemble-averaged alignment of molecular axes. As the latter rotate with respect to the probe polarization, the \LDlif{} signal becomes modulated at the frequency of molecular rotation, offering the direct measure of rotational coherence.

\begin{figure}[t]
  \includegraphics[width=.99\columnwidth]{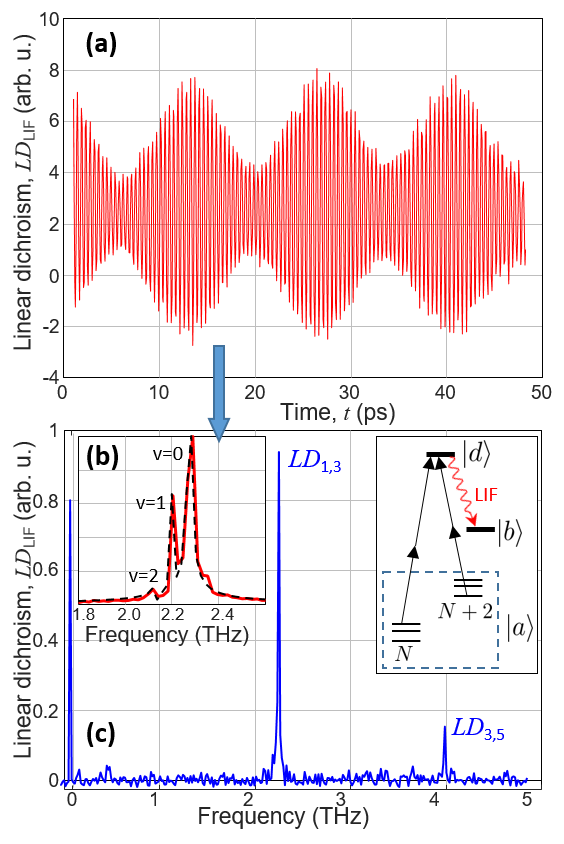}\\
  \caption{(\textbf{a}) Linear dichroism of the laser-induced fluorescence from rotationally excited \Hetwo{} molecules in the ground metastable $a$ state at $T=\SI{1.36}{K}$, as a function of the time delay between the fs kick and probe pulses. (\textbf{b}) Fourier transform of the trace in panel (\textbf{a}) showing the ro-vibrational splitting of the $LD_\textsc{1,3}$ line (solid red curve) with labeled vibrational branches, and a fit to the known gas-phase spectrum (dashed black curve). (\textbf{c}) Fourier transform of a different \LDlif{(t)} signal (shorter delay scan with a finer step) with two rotational peaks, $LD_{1,3}$ and $LD_{3,5}$.  Right diagram: relevant energy level scheme with the interfering two-photon absorption pathways and the observable LIF channel.}
  \label{fig-LIF_LD}
\end{figure}
An example of the \LDlif{(t)} signal, recorded as a function of the kick-probe delay is shown in Fig.~\ref{fig-LIF_LD}(\textbf{a}). The main oscillation frequency of $\SI{2.28(2)}{THz}$ corresponds to the energy difference $\Delta E_{1,3}/h=\SI{2.27}{THz}$ between the $N=1$ and $N=3$ rotational states of the ground vibrational level ($v=0$) of the \Astate{} manifold. The observed oscillations are the result of the quantum coherence between the $N=1$ and $N=3$ states induced by the kick pulse (hence, labeled as $LD_{1,3}$). Owing to this coherence, the two-photon $a\rightarrow d$ absorption channels originated from these two states and sharing the same rotational level in the upper \Dstate{} manifold, interfere as schematically illustrated by the diagram in Fig.~\ref{fig-LIF_LD}. The interference leads to the time-dependent total absorption, and hence the $d\rightarrow b$ fluorescence intensity, oscillating at the frequency $\nu _{1,3}=\Delta E_{1,3}/h$.

The apparent slow amplitude modulation in Fig.~\ref{fig-LIF_LD}(\textbf{a}) is due to the frequency beating between multiple vibrational states with slightly different rotational constants. The Fourier transform of the \LDlif{(t)} signal is plotted in the inset to Fig.~\ref{fig-LIF_LD}(\textbf{b}), showing the ro-vibrational splitting of the $LD_{1,3}$ rotational line. Since the frequency bandwidth of our pulses ($\approx \SI{14}{THz}$ FWHM) is smaller than the excimer's vibrational frequency ($\SI{54}{THz}$ \cite{NIST}), the vibrational excitation is inherent in the energetic process of the pump-induced \Hetwo{} formation \cite{Keto1972}. Clearly, the vibrational relaxation is far from complete \SI{1}{ms} after the pump pulse (at the arrival time of the kick-probe pulse pair), in agreement with the previously determined vibrational decay time of order of \SI{100}{ms} \cite{Eltsov1995}. Applying the known gas-phase molecular parameters \cite{Focsa1998, Semeria2018} results in a good fit of the observed ro-vibrational spectrum (dashed black curve), indicating that within the experimental uncertainty of $\approx\SI{10}{GHz}$, the rotational constants in the three vibrational states are unaffected by the liquid environment.

Fourier transform of a delay scan with a lower frequency resolution but higher frequency range reveals the second excited rotational line in the \LDlif{} spectrum, corresponding to the laser-induced coherence between the $N=3$ and $N=5$ rotational levels [$LD_{3,5}$ in Fig.~\ref{fig-LIF_LD}(\textbf{c})]. Similar to $LD_{1,3}$, the frequency of the second rotational peak $\nu_{3,5}=\SI{4.10(2)}{THz}$ agrees well with the energy difference between the $N=5$ and $N=3$ rotational levels of the $a$ state in the gas phase (\SI{4.08}{THz}).

\begin{figure}[t]
  \includegraphics[width=.9\columnwidth]{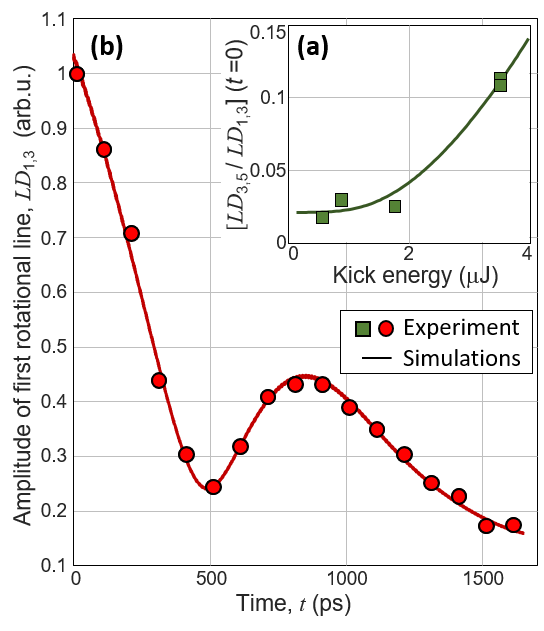}\\
  \caption{(\textbf{a}) Amplitude ratio of the two rotational lines in the linear dichroism spectrum, $LD_{3,5}$ and $LD_{1,3}$, as a function of the kick pulse energy (green squares) at $T=\SI{1.35}{K}$. (\textbf{b}) Long-time amplitude dependence of the first rotational peak at $T=\SI{1.95}{K}$, normalized to $LD_{1,3}(t=0)$ (red circles). In both panels, solid lines represent the results of numerical simulations (see text for details).}
  \label{fig-LD_time}
\end{figure}

Unlike the case of vibrational excitation, transferring the rotational population from the ground $N=1$ to the excited $N=3$ and $N=5$ states requires two-photon Raman frequencies well within the bandwidth of our kick pulses. One may ask then whether the $LD$ lines originate from the rotationally hot molecules created by the pump pulse, which have not decayed yet to the ground rotational state, or whether they stem from the molecules coherently excited by the kick pulse. To answer this question, we measured the ratio of the second-to-first rotational peak amplitudes, $LD_{3,5}/LD_{1,3}$, as a function of the pulse energy. The results are shown by green squares in Fig.~\ref{fig-LD_time}(\textbf{a}). The quick drop in the relative amplitude of the second peak with decreasing pulse intensity indicates the degree of rotational excitation largely controlled by the kick pulse.

To further support this conclusion, we carried out numerical calculations of the expected ratio between the two $LD$ peaks by solving the Schr\"{o}dinger equation in the rigid-rotor approximation (see Supplemental Material for details \cite{Supplement}). In Fig.~\ref{fig-LD_time}(\textbf{a}) we plot the ratio $LD_{3,5}/LD_{1,3}$ calculated for the experimentally used kick energies. The fit provides us with the rotational population of $N=3$ and $N=5$ levels prior to the arrival of the kick pulse, which are respectively 0.5\% and 0.05\% (upper confidence limits of ~5\% and ~0.2\%). This suggests that the majority of \Hetwo{} dimers have relaxed to the ground rotational $N=1$ state \SI{1}{ms} after their creation by the pump pulse, indicating a rotational decay constant much shorter than that found in earlier studies ($\approx \SI{15}{ms}$ \cite{Eltsov1998}). The numerical calculations also show the major re-distribution of, and hence the possibility to control, the rotational population by the kick pulse. With the energy of the latter at \SI{3.5}{\mu J} ($\approx \SI{5e11}{W/cm^2}$), more than 15\% of molecules are occupying $N=3$, and almost 2\% are at $N=5$ [in thermal equilibrium, the former (latter) would correspond to rotational temperatures of \SI{43}{K} (\SI{64}{K})].

While the analysis presented in Fig.~\ref{fig-LD_time}(\textbf{a}) offers a method of determining the decay of rotational \textit{population}, our approach also provides a way of measuring the decay of rotational \textit{coherence}. Both peaks in the linear dichroism spectrum exhibit strong time dependence, shown for $LD_{1,3}(t)$ in Figs.~\ref{fig-LD_time}(\textbf{b}). Here, a fine scan from $t$ to $(t+\SI{20}{ps})$ has been carried out for calculating the amplitude of the $LD_{1,3}$ peak at each (coarse) value of $t$ between 0 and \SI{1.65}{ns}. The oscillatory behavior is a consequence of the spin-rotational and spin-spin interactions, which split each rotational $N$-state into three $J=\left\{N,N\pm1\right\}$ states \cite{Lichten1974} (see the level diagram in Fig.~\ref{fig-LIF_LD}).

To verify this conclusion, we modeled the expected signal numerically as
\begin{equation}\label{eq-LDsignal}
LD_{1,3}(t)=\sum_{k=1..5} c^k_{1,3} \cos(2\pi \nu^k_{1,3} t) \times e^{-t/\tau_{1,3}},
\end{equation}
where $\nu ^k_{1,3}$ are the frequencies of the five transitions allowed by the selection rules (see Supplemental Material for details \cite{Supplement}), and calculated using the known accurate values for the spin-rotational and spin-spin coupling strength in the ground state of \Hetwo{} \cite{Focsa1998, Semeria2018}. Being on the scale of $\approx\SI{2}{GHz}$ , the splitting is significantly smaller than the kick bandwidth, justifying our assumption that all coherences are created with the same phase. On the other hand, coefficients $c^k_{1,3}$ account for the differences in the two-photon $J$-dependent matrix elements between different absorption pathways. Here, we used these coefficients as free fitting parameters, leaving the comparison to their \textit{ab initio} values to future theoretical analysis.

Our assumption of a single decay constant $\tau _{1,3}$ in Eq.~\ref{eq-LDsignal} is justified by the quality of the fit in Fig.~\ref{fig-LD_time}(\textbf{b}). From the fit, we extract the coherence lifetime $\tau_{1,3}=\SI{1.0(5)}{ns}$, during which the molecule completes more than a thousand full rotations. The corresponding rotational linewidth of $\approx \SI{0.3}{GHz}$ is significantly narrower than the scan-length limited lines in Figs.~\ref{fig-LIF_LD}(\textbf{b,c}). We note that $v>0$ vibrational branches, not included in the fit, add fast oscillations around the plotted curve without changing the optimal fit parameters. At this time, we were unable to apply the same numerical analysis to the much weaker second rotational line ($LD_{3,5}$). Improving the signal quality and comparing the two decays is the objective of current investigation.

\begin{figure}[t]
  \includegraphics[width=.99\columnwidth]{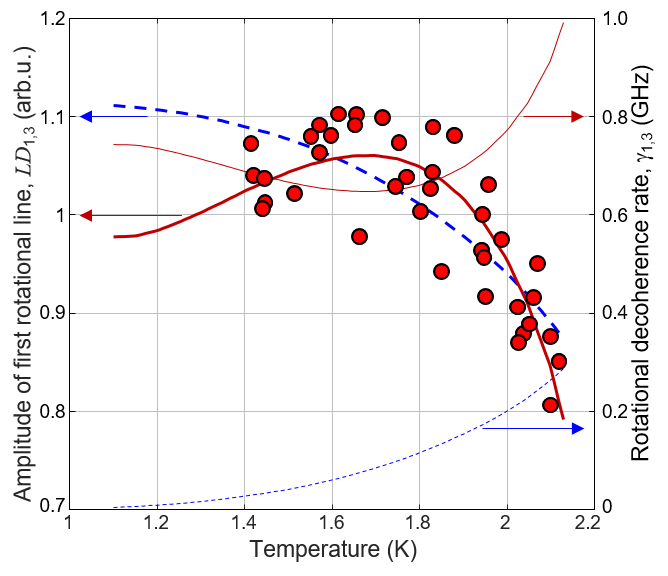}\\
  \caption{Dependence of the amplitude of the first rotational peak ($LD_{1,3}$) on the temperature of the superfluid at the kick-probe delay of $t=\SI{850}{ps}$ (red circles, left vertical axis). Thick dashed blue and solid red lines are fits to the equilibrium kinematic decoherence model [Eq.~(\ref{eq-decayRotons2})] and its non-equilibrium modification [Eq.~(\ref{eq-u2Pulse})], respectively. Thin dashed blue and solid red lines are corresponding decoherence rates (right vertical axis).}
  \label{fig-LD_temperature}
\end{figure}

One of the main advantages of studying molecular dynamics in bulk liquid helium is the ability to vary the temperature and pressure of the superfluid, probing the macroscopic nature of superfluidity. Here, we explored the temperature dependence of the rotational coherence between $N=1$ and $N=3$ rotational levels, reflected by the amplitude of the $LD_{1,3}$ peak in the dichroism spectrum. The experimental result, measured at a fixed kick-probe delay of \SI{850}{ps}, is shown by red circles in Fig.~\ref{fig-LD_temperature}. A clear decrease of $LD_{1,3}$ with temperature increasing towards the lambda point is a signature of the apparent interaction between the liquid and the laser-induced coherent rotation of helium dimers.

Unlike the $T$-dependent change in the total fluorescence signal \cite{Supplement}, the observed rotational decoherence cannot be attributed to bimolecular collisions. Indeed, from the known diffusion constant of the \Hetwo{} molecules in our temperature range ($\lesssim\SI[parse-numbers=false]{10^{-3}}{cm^2s^{-1}}$ \cite{McKinsey2005}), their average displacement on the time scale of our experiment is about \SI{10}{nm}, which is significantly smaller than the inter-molecular separation of $>\SI{300}{nm}$ for the experimentally determined molecular density of \SI{2e13}{cm^{-3}} (see Supplemental Material for details \cite{Supplement}).

On the other hand, scattering of thermal quasiparticles (i.e. the normal component of the liquid) on molecular rotors could be responsible for the observed temperature dependence of \LDlif{}.  In a simple kinematic picture, where the (quasi)stationary dimers are colliding with He atoms moving with the velocity of first sound $u_1(T)$, one can write:
\begin{eqnarray}
\hspace{-5mm} LD_{1,3}(t,T)&=& LD^{(0)}_{1,3} \times \exp\left[-\gamma_{1,3}(T)\,t \right] \label{eq-decayRotons1} \\ &=&LD^{(0)}_{1,3} \times \exp \left[-N^{eq}_{n}(T)\,\sigma_{1,3}(T)\,u_1(T)\,t\right], \label{eq-decayRotons2}
\end{eqnarray}
where $\gamma_{1,3}$ is the decoherence rate, $N^{eq}_n$ is the equilibrium atom number density of the normal fluid, and $\sigma_{1,3}$ is the scattering cross-section. Given the unknown $T$ dependence of $\sigma_{1,3}$, we used it as a temperature-independent single fitting parameter. The best fit, shown with the thick dashed blue curve in Fig.~\ref{fig-LD_temperature}, captures the overall trend in the data, but fails to reproduce the non-exponential flattening of the curve at lower temperatures. It also results in the decoherence rate (thin dashed blue curve), which is significantly lower than $\gamma_{1,3} \approx \SI{1}{GHz}$ observed in our scans of the kick-probe delay at $T=\SI{1.95}{K}$ discussed earlier (Fig.\ref{fig-LD_time}).

While further theoretical investigation of $\sigma_{1,3} (T)$ may help reconcile our data with the collisional model of Eq.~(\ref{eq-decayRotons2}), we note that it is based on the assumption of a thermal equilibrium between molecular rotors and the surrounding liquid. This assumption justified the use of the time-independent normal fraction $N^{eq}_n(T)$. However, the impulsive excitation of \Hetwo{} by intense kick pulses may also create a non-equilibrium state, in which molecular rotors find themselves surrounded by a microscopic local volume of ``hot'' liquid. The sudden imbalance of entropy may trigger a coherent pulse of second sound, initiating a flow of the normal component away from the molecule, and a corresponding counterflow of superfluid towards it. Consider for simplicity a Gaussian pulse of width $w$, traveling with the speed of second sound $u_2(T)$, and describing the non-equilibrium density of the normal fraction at the location of the molecule:
\begin{equation}\label{eq-u2Pulse}
N^{neq}_n(T,t)=N\times \exp \left[- \bigl( u_2(T)\, t \bigr)^2/ w^2 \right],
\end{equation}
where $N$ is the total density of the liquid -- all of it in the normal phase at time zero. Substituting $N^{eq}_n(T)$ in Eq.~(\ref{eq-decayRotons2}) by this time-dependent $N^{neq}_n(T,t)$, and fitting it to the data in Fig.~\ref{fig-LD_temperature} using $\sigma_{1,3}$ and $w$ as free parameters, results in the thick solid red line. In contrast to the equilibrium model, the decoherence rate (thin solid red line) mediated by the wave of second sound, is much more consistent with our findings from the delay scan at \SI{1.95}{K}.

The non-equilibrium picture also better reproduces the flattening of the $LD_{1,3}(T)$ data between 1.4 and \SI{1.8}{K}. The local minimum of $\gamma _{1,3}$ in this temperature window stems from the corresponding local maximum in the speed of second sound \cite{Donnelly2009}. The faster the entropy pulse, the faster the counterflow of the frictionless superfluid component towards the molecular rotor, the slower its rotational decoherence. As the speed of second sound decreases, with $T$ increasing beyond \SI{1.8}{K}, the heat wave created by the kick pulse travels shorter distance away from the molecular rotor in a given amount of time. The correspondingly slower influx of the superfluid component results in faster decoherence and a lower signal amplitude. We note that thermal diffusion, which becomes faster with increasing $T$, would result in the opposite dependence of our signal on temperature, thus making it inadequate for explaining our experimental findings.

The decoherence cross-section $\sigma_{1,3}=\SI{2.5e-2}{\AA^2}$, extracted from the fit to the non-equilibrium model, appears to be four orders of magnitude smaller than the size of the \Hetwo{} bubble. This indicates very weak coupling between the normal fluid and the spherical \Astate{} state, and explains why, within our experimental uncertainty, the rotational constants of the excimer in different vibrational and electronic spin states seem to be unaffected by the surrounding superfluid. The \SI{22}{nm} width of the second sound pulse, provided by the fit, is larger than the distance of \SI{17}{nm} covered by the pulse in \SI{850}{ps}, which justifies the proposed far-from-equilibrium scenario. The latter could also explain the relatively large scatter of experimental data in Fig.~\ref{fig-LD_temperature}, which we could not trace to any source of instrumental noise.

In summary, we report the first experimental observation of the laser-induced coherent molecular rotation in bulk superfluid liquid helium. Our time-resolved method enables us to detect and study various rotational dynamics in three different time windows: (i) we characterize the degree of rotational cooling on the millisecond timescale; (ii) probe the ro-vibrational, spin-rotational and spin-spin dynamics on the picosecond timescale; and (iii) investigate the decay of rotational coherence on the nanosecond timescale.

By measuring the temperature dependence of the coherent rotational signal, we identify two possible decoherence mechanisms of qualitatively different nature: one mediated by the normal component of the helium bath in thermal equilibrium with the rotating molecule, and another one based on non-equilibrium dynamics of the superfluid, governed by the wave of second sound. We note that such non-equilibrium response of \HeII{} to the sudden injection of energy by an intense ultrashort laser pulse has recently been observed in the ultrafast dynamics of rotons \cite{Milner2023a}. Since rotons are predominant collective excitations of the normal component at $T\gtrsim \SI{1}{K}$ \cite{McKinsey2005}, one may also expect the latter to interact with a suddenly initiated rotation of a molecular probe in a non-equilibrium fashion. Work is underway to further investigate the molecule-superfluid interaction under variable temperature and pressure, to better differentiate between the proposed equilibrium and non-equilibrium models.

We also demonstrate the ability to vary the degree of rotational excitation, which offers a method of measuring the anisotropic polarizability of the helium dimer. Finally, new information about the rotational relaxation of \Hetwo{} in \HeII{} may also help improve the methods of LIF-based molecular tagging \cite{McKinsey2005, Rellergert2008, Guo2009} in the studies of the counterflow \cite{Guo2010} and quantum turbulence \cite{Zmeev2013, Guo2014, Gao2015} in superfluids, as well as open new avenues for studying the microscopic implications of superfluidity with molecular nano-probes.

\begin{figure*}[t]
  \includegraphics[width=.9\textwidth]{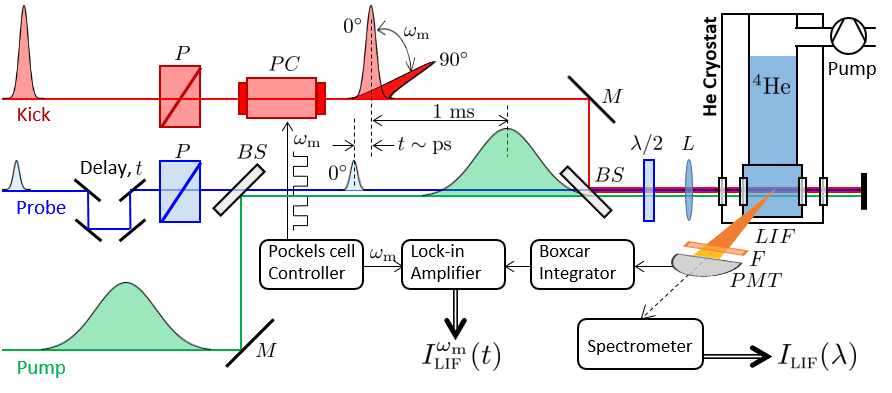}\\
  \caption{Scheme of the experimental setup. Pump (lower green), kick (upper red) and probe (middle blue) pulses (all at $\lambda \approx \SI{800}{nm}$) are focused at the center of the liquid helium cryostat. Polarization of the kick pulses is modulated with a Pockels cell ($PC$) at the frequency $\omega _{m}$. Laser-induced fluorescence ($LIF$) is collected in the direction orthogonal to the laser beams, filtered around \SI{640}{nm}, and sent to either a spectrometer, or a photo-multiplier tube ($PMT$). The PMT signal is gated with a boxcar integrator around the arrival time of probe pulses, and fed to the lock-in amplifier, which amplifies the signal at $\omega _{m}$. $P$: polarizer, $M$: mirror, $BS$: beam splitter, $\lambda /2$: half-wave plate, $L$: lens ($f=\SI{25}{cm}$), $F$: bandpass filter (\SI{20}{nm} FWHM).}
  \label{fig-Setup}
\end{figure*}
\section*{Acknowledgments}
We would like to thank Dr. Jussi Eloranta and Dr. Wei Guo for many helpful discussions.


%

\section*{Supplemental Material}

\subsection*{Generation of \Hetwo{} excimers solvated in \HeII{}}
Our experiments are performed in a custom-built helium cryostat (Fig.~\ref{fig-Setup}). By pumping on the helium, the temperature of the liquid can be varied between $\approx\SI{1.4}{K}$ and $\SI{4.2}{K}$, while the pressure above the surface is at the saturated vapor pressure (SVP). Three laser pulses -- pump, kick and probe -- are delivered to the cryostat at the repetition rate of \SI{1}{KHz}, and are focused in LHe with a \SI{250}{mm}-focal length lens. Extracted from the same ultrafast Ti:Sapph laser system, they share the same central wavelength of $\SI{798}{nm}$ and bandwidth of $\SI{30}{nm}$ (full width at half maximum, FWHM), but differ in pulse length, energy, and the time of arrival. The kick-probe pulse pair is delayed from the pump by $\approx\SI{1}{ms}$, whereas the delay within the pair can be scanned up to \SI{1.2}{ns} with fs accuracy.

\begin{figure}[t]
  \includegraphics[width=.9\columnwidth]{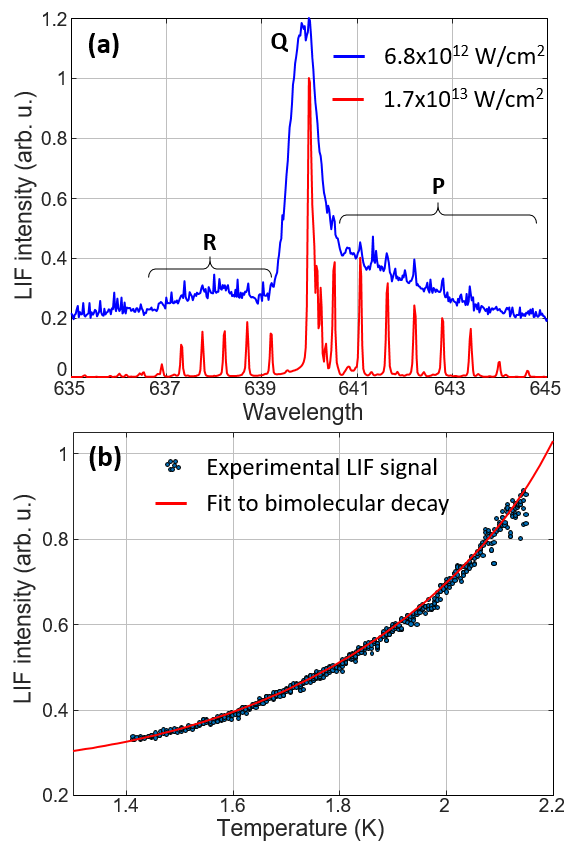}\\
  \caption{(\textbf{a}) Pump-induced $d \rightarrow b$ fluorescence spectrum of \Hetwo{} at $T=\SI{2}{K}$ with P, Q and R rotational branches labeled. The broadening of rotational lines with decreasing pump intensity demonstrates the transition between the molecules in macroscopic gas pockets (``gas phase'', lower red line) and solvated molecules (``bubble phase'', upper blue line), and illustrates the effect of the superfluid on the molecular rotation. (\textbf{b}) Intensity of the probe-induced fluorescence as a function of temperature (dots) and its fit to the expected bimolecular decay (solid red line, see text for details).}
  \label{fig-LIF}
\end{figure}

Pump pulses, stretched to $\approx \SI{2}{ps}$ and carrying $\SI{80}{\mu J}$ per pulse, are used to create helium excimers. Their peak intensity of \SI{4e11}{W/cm^2} is significantly below the breakdown threshold $I_\text{break}\approx\SI{5e13}{W/cm^2}$, determined in previous studies \cite{Benderskii1999, Gao2015}. We note, however, that sub-breakdown intensities do not guarantee the production of the desired ``bubble phase'', i.e. an ensemble of isolated \Hetwo{} molecules, each solvated in the liquid in its own bubble. To illustrate this, Fig.~\ref{fig-LIF}(\textbf{a}) shows the observed fluorescence spectrum corresponding to pump intensities of \SI{1.7e13}{W/cm^2} (lower red curve) and \SI{6.8e12}{W/cm^2} (upper blue curve). Even though both intensities are below $I_\text{break}$, the gas-phase-like narrow rotational lines in the lower trace indicate that the molecules are created in macroscopic \textit{gas pockets} \cite{Mendoza2016}. We found the transition between the bubble and gas phases to occur at $I_\text{gas}\approx \SI{5e12}{W/cm^2}$, which dictated our choice of all pulse intensities well below this threshold value.

To further verify the important aspect of preparing the excimers in the solvated bubble state, we investigated the influence of the LHe temperature on the total LIF signal. The latter is proportional to the time-dependent \Hetwo{} number density $N(t)$. The time dependence is governed by the bimolecular annihilation reaction, $\mathrm{He}_2^* + \mathrm{He}_2^* \rightarrow \mathrm{He}_2^{**} + 2\mathrm{He}$ \cite{Keto1974, Benderskii1999}:
\begin{equation}\label{eq-density}
    N(t)=N_0/\left[1+K(T) N_0 t\right].
\end{equation}
The dependence on temperature enters through the reaction rate $K(T)$. In the bubble phase, $K(T)$ is determined by the diffusion of \Hetwo{} molecules in the liquid due to their scattering on thermal rotons \cite{Keto1972, Hereford1972}. As the roton energy $\Delta (T)$ decreases with increasing temperature \cite{Andersen1996}, their equilibrium density grows proportionally to $\sqrt{T} \exp\left[ -\Delta (T)/T \right]$ \cite{Bedell1982}, causing the scattering length and the diffusion coefficient to decrease. Slower diffusion at higher $T$ results in a lower annihilation rate, and correspondingly larger number of excimers at any given time.

Dark dots in panel (\textbf{b}) of Fig.~\ref{fig-LIF} show our measured temperature dependence of the fluorescence intensity, induced by a fs probe pulse (pulse length of \SI{70}{fs} FWHM, intensity of \SI{1.3e12}{W/cm^2}) following the ps pump pulse after a fixed delay of $\approx \SI{1}{ms}$. The red solid line is a fit to Eq.~(\ref{eq-density}), with the known roton energy $\Delta (T)$ from the neutron scattering experiments \cite{Andersen1996}. An excellent fit to the roton-mediated diffusion model confirms the production of \Hetwo{} molecules properly solvated in liquid helium. Note that in the gas phase, one would expect a very different rate of bimolecular decay, owing to the increasing gas pressure, and hence higher annihilation rate, with increasing temperature. From the fit in Fig.~\ref{fig-LIF}(\textbf{b}), the initial molecular number density ($N_{0}$, the only fitting parameter) is \SI{1.9(1)e13}{cm^{-3}}.

\subsection*{Rotational excitation of \Hetwo{} excimers}
To excite the rotation of helium excimers we send the linearly polarized femtosecond kick pulse prior to the probe (upper red in Fig.~\ref{fig-Setup}). Utilizing the \SI{1}{KHz} repetition rate of our laser system, we make the kick-probe pair trail the pump pulse by about \SI{1}{ms}. At that time, the amount of \Hetwo{} molecules which survived the bimolecular reaction is still quite high, yet given the experimentally observed decay of the pump-induced fluorescence on the timescale of a few tens of nanoseconds \footnote{This observation is consistent with the earlier study of Benderskii et. al \cite{Benderskii1999}}, all these molecules have already decayed to their lowest metastable electronic state \Astate{}. On the other hand, the rotational relaxation is expected to occur on the scale of a few milliseconds \cite{Eltsov1998}. As we demonstrate in this work, even after \SI{1}{ms} most of the molecules have also relaxed to the ground rotational state corresponding to the angular momentum (excluding electronic spin) $N=1$ (due to the nuclear spin statistics, only odd values of $N$ are allowed in \Astate{} \cite{Focsa1998}).

\subsection*{Detection of \Hetwo{} rotation}
Our method of detecting molecular rotation is based on the anisotropic absorption cross-section, common to linear molecules \cite{Auzinsh1995}. Two probe photons with a wavelength of \SI{800}{nm} promote the excimer from the ground $a$ to the excited $d$ state \cite{Benderskii2002, Rellergert2008} with an absorption rate dependent on the angle between the molecular axis and the vector of the probe polarization. The difference in the absorption of two orthogonally polarized probe pulses (known as ``linear dichroism'', $LD$) corresponds to the anisotropy of the ensemble-averaged distribution of molecular axes, whereas its time dependence reflects the rotational dynamics of the molecules \footnote{The approach is similar to the polarization-based studies of the laser-induced rotation of gas-phase molecules \cite{Renard2003}}. Since we detect this linear dichroism via the induced fluorescence on the $d\rightarrow b$ transition, we refer to it as \LDlif{}.

We keep the probe polarization constant and modulate the polarization direction of the kick pulses between $0^{\circ}$ and $90^{\circ}$ with a Pockels cell ($PC$ in Fig.~\ref{fig-Setup}). The \LDlif{} signal is defined as:
\begin{equation}\label{eq-LD_LIF}
LD_\textsc{lif}=\frac{I_\textsc{lif}^\parallel - I_\textsc{lif}^\perp}{(I_\textsc{lif}^\parallel + I_\textsc{lif}^\perp)/2},
\end{equation}
where $I_\textsc{lif}^{\parallel,\perp}$ is the fluorescence intensity recorded (by a photo-multiplier tube) with the kick polarization respectively parallel or perpendicular to the fixed probe polarization. We use a Boxcar integrator to gate the fluorescence signal around the arrival time of the kick-probe pulse pair. A lock-in amplifier is employed to retrieve the dichroism  from the LIF intensity as the signal component at the polarization modulation frequency $\omega _{m} \approx\SI{200}{Hz}$. To eliminate possible instrumental artifacts due to our detection geometry, we use a half-wave plate ($\lambda /2$ in Fig.~\ref{fig-Setup}) to rotate all  polarization vectors by $45^{\circ}$ with respect to the excitation-observation plane. The angle of the plate is fine-tuned to bring the \LDlif{} signal to zero when the probe pulse is blocked. We note that due to the undesired \textit{kick}-induced fluorescence, inadvertently entering the denominator in Eq.~\ref{eq-LD_LIF}, we are currently unable to extract reliable absolute values of \LDlif{}.

\subsection*{Calculating the effect of a fs rotational kick}
Consider a rotational state $\psi_{J,M}(t)=\sum_{J,M} c_{J,M} \exp(-iE_J t) \ket{J,M}$ interacting with a laser field $\E_\text{kick}(t)$. Here, $J$ and $M$ are the molecular total angular momentum (including electronic spin) and its projection on the vector of kick polarization, whereas $c_{J,M}$ and $E_J$ are the amplitude and energy of the corresponding eigenstate. The interaction potential is given by \cite{Floss2012}
\begin{equation}\label{eq-potential}
V(t)=-\frac{1}{4}\Delta\alpha \cos^2(\theta) \E_\text{kick}(t),
\end{equation}
where $\Delta \alpha =\SI{35.1}{\AA^3}$ is the difference between the molecular polarizability along and perpendicular to the molecular axis \footnote{J.~Eloranta, private communication. Calculated using the method of Coupled Cluster with Single and Double substitutions, and the basis set from Ref.~\citenum{Eloranta2001}}, and $\theta $ is the angle between the molecular axis and the laser field polarization. We numerically solved the Schr\"{o}dinger equation in the rigid-rotor approximation, assuming that the kick length is much shorter than the period of the molecular rotation. For simplicity, here we also neglect the effect of the electronic spin (which is discussed in the next section) and take $J\equiv N$ and $E_J\equiv E_N$.

Linearly polarized kick field leaves the molecule in a coherent superposition of states with $\Delta N=\pm2$ and $\Delta M=0$. The numerical solution of the Schr\"{o}dinger equation provides us with the complex amplitudes $c_{N,M}$ of those states right after the kick. The observed $LD_{N,N+2}$ signal, oscillating at the frequency $\Delta E_{N,N+2}$, is proportional to the real part of the product $c_{N,M}c^*_{N+2,M}$, summed over all independent $M$ channels.

\subsection*{Spin-rotational/spin-spin splitting of rotational lines}
Consider the first rotational line $LD_{1,3}$, corresponding to the coherent superposition of $N=1$ (split into $J_1=\left\{0,1,2\right\}$) and $N=3$ (split into $J_3=\left\{2,3,4\right\}$), created by the kick pulse. The absorption of two probe photons on the $a\rightarrow d$ transition must obey the selection rule $\Delta J=0,2$. Therefore, there are five $(J_1,J_3=\left\{J_1,J_1\pm2,J_1\pm4\right\})$ pairs producing the $LD$ signal at the approximate frequency $\nu^k_{1,3}\approx\SI{2.27}{THz}$: $(0,2),(0,4),(1,3),(2,2)$ and $(2,4)$. Beating of these five frequencies results in the observed oscillations of $LD_{1,3}$ with a minimum around \SI{500}{ps} (see Fig.~\ref{fig-LD_time}).

\end{document}